\documentclass[aps,showpacs,pra,superscriptaddress,]{revtex4}
\usepackage{amsmath}
\usepackage{amsfonts}
\usepackage{amssymb}
\usepackage{bm}
\usepackage{graphicx}

\begin{document}

\title{Novel solitary and periodic waves in quadratic-cubic
non-centrosymmetric waveguides}
\author{Houria Triki}
\affiliation{Radiation Physics Laboratory, Department of Physics, Faculty of Sciences,
Badji Mokhtar University, P. O. Box 12, 23000 Annaba, Algeria }
\author{ Vladimir I. Kruglov}
\affiliation{Centre for Engineering Quantum Systems, School of Mathematics and Physics,
The University of Queensland, Brisbane, Queensland 4072, Australia}

\begin{abstract}
We present a wide class of novel solitary and periodic waves in a
non-centrosymmetric waveguide exhibiting second- and third-order
nonlinearities. We show the existence of bright, gray, and W-shaped solitary
waves as well as periodic waves for extended nonlinear Schr\"{o}dinger
equation with quadratic and cubic nonlinearities. We also obtained the exact
analytical algebraic-type solitary waves of the governing equation,
including bright and W-shaped waves. The results illustrate the propagation
of potentially rich set of nonlinear structures through the optical
waveguiding media. Such privileged waveforms characteristically exist due to
a balance among diffraction, quadratic and cubic nonlinearities.
\end{abstract}

\pacs{05.45.Yv, 42.65.Tg}
\maketitle
\affiliation{$^{1}${\small Radiation Physics Laboratory, Department of Physics, Faculty
of Sciences, Badji Mokhtar University, P. O. Box 12, 23000 Annaba, Algeria}\\
$^{2}${\small Centre for Engineering Quantum Systems, School of Mathematics
and Physics, The University of Queensland, Brisbane, Queensland 4072,
Australia}}

\section{Introduction}

Interest in nonlinear localized waves also\ called solitons has grown
considerably in recent times due to their appearance in various physical
systems. The occurrence of such structures range from fluid dynamics \cite%
{Kodama10}, Bose-Einstein condensates \cite{Burger}-\cite{Strecker},
fiber-optic communications and photonics in general \cite{L,Y}, to nuclear
physics \cite{Arriola}, and plasmas physics \cite{Infeld,Shukla}. The
research results have shown that there exist two distinct functional forms
of localized waves, hyperbolic and algebraic forms, which play the same role
in the wave dynamics. It is worthwhile to mention here that algebraic-type
solitary waves are localized more weakly compared with the conventional
(hyperbolic) solitons \cite{Hayata}.

Theoretically, the study of propagation properties of localized waves in a
Kerr dielectric guide involves solving the nonlinear Schr\"{o}dinger (NLS)
equation that includes the group velocity dispersion and self-phase
modulation \cite{Kodama}. Such underlying model has also been applied to the
description of matter waves in Bose-Einstein condensates \cite{Beitia}. In
the latter setting, the equation is usually called the Gross- Pitaevskii
equation (GPE) \cite{Pitaevskii}. We mention in passing that the NLS model
is also relevant for electromagnetic pulse propagation in negative index
materials \cite{Scalora}.

Recent advances in the study of optical materials have demonstrated that the
application of the NLS model for a more realistic description of wave
dynamics in many practical materials imposes the inclusion of additional
nonlinear and dispersive terms in the underlying equation \cite{Avelar}-\cite%
{Nikola}. In this context, several generalizations of the NLS equation with
different forms of nonlinearities have been developed to study the wave
evolution in diverse physical systems, including cubic-quintic \cite{Avelar}%
, cubic-quintic-septimal \cite{R,TH}, polynomial \cite{Nikola}, and
saturable \cite{Vasantha} nonlinearity. The quadratic-cubic NLS equation is
a newly extension of the NLS equation which has gathered significant
attention in recent years. Such equation may be used as an approximate form
of the GPE for quasi-one-dimensional Bose-Einstein condensate with contact
repulsion and dipole-dipole attraction \cite{Fujioka}. This model can be
also applied for the description of light beam propagation in a
non-centrosymmetric waveguide exhibiting second- and third-order
nonlinearity \cite{Pal}. Due to its physical importance, this nonlinear wave
evolution equation has been analyzed from different points of view. For
instance, Cardoso et al. have obtained the localized solutions for
inhomogeneous quadratic-cubic NLS equation and studied their stability with
respect to small random perturbations \cite{Cardoso}. Triki et al. have
analyzed this equation with space and time modulated nonlinearities in
presence of external potentials in the context of Bose-Einstein condensates 
\cite{TH0}. In \cite{TH1}, the soliton solutions and the conservation laws
of the equation were reported. In \cite{Pal}, the chirped self-similar wave
solutions of the equation were constructed by employing the similarity
transformation method. Some soliton solutions of the equation have been also
obtained by means of the extended trial equation method in \cite{TH2}.

Looking for more novel solutions to quadratic-cubic NLS model is an
important direction in the studies of nonlinear matter and optical wave
propagations. In particular, obtaining new solutions in analytic form will
be a remarkable contribution to well understand physical phenomena in
various dynamical systems where the quadratic-cubic NLS equation can provide
a realistic description of the waves. In this paper, we present new types of
exact analytical localized and periodic wave solutions for the
quadratic-cubic NLS equation.

The paper is organized as follows. In Sec. II, we present the
quadratic-cubic NLS model describing the propagation of light beams in
non-centrosymmetric waveguides, and we give a detailed study of families of
novel solitary and periodic wave solutions of the model. We also examine
here the existence condition for algebraic solitary waves of the underlying
equation. Finally, we present some conclusions in Sec. III.

\section{Model and novel solitary and periodic wave solutions}

The propagation of light beams in quadratic-cubic non-centrosymmetric
waveguides is modeled by the following quadratic-cubic NLS equation \cite%
{TH1,Pal,TH2}, 
\begin{equation}
i\frac{\partial E}{\partial z}-\frac{\alpha }{2}\frac{\partial ^{2}E}{%
\partial x^{2}}+\sigma |E|E+\nu |E|^{2}E=0,  \label{1}
\end{equation}%
where $z$ is the longitudinal variable representing propagation distance, $x$
is transverse variable, and $E(z,x)$ is the complex envelope of the
electrical field. The parameters $\alpha ,$ $\sigma $ and $\nu $ represent
the diffraction, quadratic and cubic nonlinearity coefficients, respectively.

To find exact solutions of Eq. (\ref{1}), we consider an ansatz solution of
the form \cite{VK}, 
\begin{equation}
E(z,x)=u(\xi )\exp [i\left( \kappa z-\Omega x+\theta \right) ],  \label{2}
\end{equation}%
where $u$ is a differentiable real function depending on the variable $\xi
=x-qz$, with $q=v^{-1}$ being the inverse velocity of the wave packet. Also, 
$\kappa $ and $\Omega $ are the respective real parameters describing the
wave number and frequency shift, while $\theta $ represents the phase of the
pulse at $z=0$.

\noindent Inserting Eq. (\ref{2}) into Eq. (\ref{1}) and separating real and
imaginary parts, we obtain,%
\begin{equation}
q=\alpha \Omega ,  \label{3}
\end{equation}

\noindent from the imaginary part, indicating that the inverse velocity $q$
is controlled by the parameters $\alpha $ and $\Omega $. The real part
yields the equation,%
\begin{equation}
-\kappa u-\frac{\alpha }{2}\frac{d^{2}u}{d\xi ^{2}}+\frac{\alpha \Omega ^{2}%
}{2}u+\sigma |u|u+\nu u^{3}=0,  \label{4}
\end{equation}

\noindent The latter can expressed in the form%
\begin{equation}
\frac{d^{2}u}{d\xi ^{2}}+au^{3}+b|u|u+cu=0,  \label{5}
\end{equation}

\noindent where the parameters $a$, $b$ and $c$ are given by%
\begin{equation}
a=-\frac{2\nu }{\alpha },\qquad b=-\frac{2\sigma }{\alpha },\qquad c=\frac{%
2\kappa -\alpha \Omega ^{2}}{\alpha }.  \label{6}
\end{equation}

Nonlinear differential equation (\ref{5}) with coexisting quadratic $|u|u$
and cubic $u^{3}$ describes the evolution dynamics of the field amplitude as
it propagates through the non-centrosymmetric waveguide. Nonlinear waveforms
propagating inside the waveguiding media can be readily obtained by solving
this nonlinear differential equation. We emphasis that the term $b|u|u$ in
Eq. (\ref{5}) has two different forms for positive and negative values of
the function $u(\xi )$: 
\begin{equation}
b|u|u=bu^{2}~~~(for~u(\xi )\geq 0),\qquad b|u|u=-bu^{2}~~~(for~u(\xi )\leq
0).  \label{7}
\end{equation}%
This feature is crucial for obtained exact solutions presented below. In the
following, we present novel exact solitary and periodic wave solutions for
Eq. (\ref{1}) obtained by substitution of closed form solutions of the
nonlinear differential equation (\ref{5}) into the ansatz solution (\ref{2}%
). To our knowledge, the gray solitary wave (\ref{16}), W-shaped solitary
waves (\ref{17}) and (\ref{18}), periodic wave solutions (\ref{23}) and (\ref%
{29}) presented below are firstly reported in this work. Such privileged
exact solutions are formed in the optical waveguiding media due to a balance
among diffraction, quadratic and cubic nonlinearities.

\begin{figure}[h]
\includegraphics[width=1\textwidth]{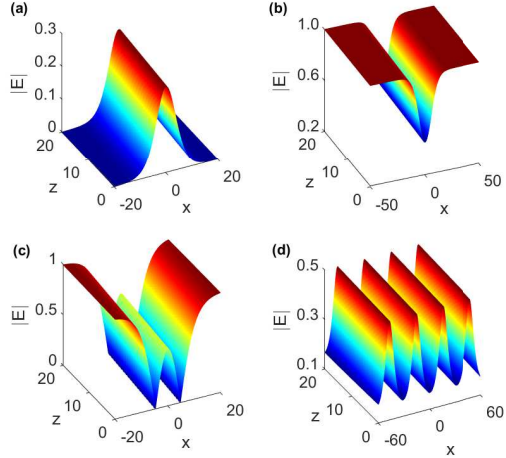}
\caption{Evolution of nonlinear wave solutions (a) bright solitary
wave (11) with parameters $\alpha =1,$ $\sigma =-0.5,$ $\nu =0.25,$ $q=0.1,$ 
$z_{0}=0,$ $\Omega =1,$ and $\kappa =0.42$ (b) gray solitary wave (16) with
parameters $\alpha =2,${\large \ }$\sigma =-0.5,${\large \ }$\nu =0.3,$%
{\large \ }$q=0.1,${\large \ }$z_{0}=0,${\large \ }$\Omega =1,$ and $\kappa
=0.8$ (c) W-shaped solitary wave (17) with parameters $\alpha =2,${\large \ }%
$\sigma =\frac{1}{15},${\large \ }$\nu =\frac{1}{45},${\large \ }$q=0.1,$%
{\large \ }$z_{0}=0,${\large \ }$\Omega =\frac{1}{3},$ and $\kappa =0.2$ (d)
periodic wave (23) with parameters $\alpha =1,$ $\sigma =0.25,$ $\nu =-0.5,$ 
$q=0.1,$ $z_{0}=0,$ $\kappa =\frac{25}{48}$ and $\Omega =1$. }
\label{FIG.1.}
\end{figure}

\begin{description}
\item[\textbf{1. Bright solitary waves}] 
\end{description}

The nonlinear differential equation (\ref{5}) supports the exact solitary
wave solution as 
\begin{equation}
u(\xi )=\frac{A}{1+D\cosh (w(\xi -\xi _{0}))}.  \label{8}
\end{equation}%
There are two different cases for parameters in Eq. (\ref{8}). In the first
case [$u(\xi )\geq 0$], the real parameters $w$, $A$ and $D$ are given by 
\begin{equation}
w=\sqrt{-c},\qquad A=-\frac{3c}{b},  \label{9}
\end{equation}%
\begin{equation}
D=\sqrt{1-\frac{9ac}{2b^{2}}},  \label{10}
\end{equation}%
where $c<0$ and $9ac/2b^{2}<1$. In the second case [$u(\xi )\leq 0$] the
parameters $w$ and $D$ have the same form, however the parameter $A$ is
given as $A=3c/b$. This is connected with relations in Eq. (\ref{7}) which
yield in this second case the replacement of parameter $b$ to $-b$. The
exact solution of Eq. (\ref{1}) for these two cases has the form,%
\begin{equation}
E(z,x)=\frac{A}{1+D\cosh (w(x-q(z-z_{0})))}\exp [i\left( \kappa z-\Omega
x+\theta \right) ].  \label{11}
\end{equation}

Figure 1(a) presents an example of propagation of the solitary wave solution
(\ref{11}) for the parameter values $\alpha =1,$ $\sigma =-0.5,$ $\nu =0.25,$
$q=0.1,$ $z_{0}=0,$ $\Omega =1,$ and $\kappa =0.42$. It is interesting to
note that the structure is a bright pulse on a zero background.

\begin{description}
\item[\textbf{2. Gray and dark solitary waves}] 
\end{description}

We have obtained the gray solitary wave solution for Eq. (\ref{5}) as
follows:%
\begin{equation}
u(\xi )=\lambda -\frac{B}{1+D\cosh (w(\xi -\xi _{0}))}.  \label{12}
\end{equation}%
Note that Eq. (\ref{12}) describes the gray solitary waves for two different
cases: (1) $u(\xi )>0$, and (2) $u(\xi )<0$. In the first case [$u(\xi )>0$]
we have the conditions as $\lambda >0$ and $\lambda -B/(1+D)>0$. In this
case the parameter $\lambda $ is given by equation $a\lambda ^{2}+b\lambda
+c=0$ which yields 
\begin{equation}
\lambda =\frac{1}{2a}\left( -b\pm \sqrt{b^{2}-4ac}\right) .  \label{13}
\end{equation}%
The parameters $w$, $B$ and $D$ are given by 
\begin{equation}
w=\sqrt{2c+b\lambda },\qquad B=-\frac{3(2c+b\lambda )}{b+3a\lambda },
\label{14}
\end{equation}%
\begin{equation}
D=\left[ 1+\frac{9a(2c+b\lambda )}{2(b+3a\lambda )^{2}}\right] ^{1/2}.
\label{15}
\end{equation}

In the second case [$u(\xi )<0$] we have the conditions as $\lambda <0$ and $%
\lambda -B/(1+D)<0$. The relations in (\ref{7}) yield in this case the
replacement of parameter $b$ to $-b$ in Eqs. (\ref{13}-\ref{15}). The exact
gray solitary wave solutions of Eq. (\ref{1}) for two cases [with $u(\xi )>0$
and $u(\xi )<0$] are given by 
\begin{equation}
E(z,x)=\left( \lambda -\frac{B}{1+D\cosh (w(x-q(z-z_{0})))}\right) \exp
[i\left( \kappa z-\Omega x+\theta \right) ].  \label{16}
\end{equation}%
We emphasis that Eq. (\ref{13}) defines two different values for parameters $%
\lambda $. Hence, the above solitary waves are determined for each value of $%
\lambda $.

In Fig. 1(b), we have plotted an example of propagation of the gray solitary
wave solution (\ref{16}) for the parameter values{\large \ }$\alpha =2,$%
{\large \ }$\sigma =-0.5,${\large \ }$\nu =0.3,${\large \ }$q=0.1,${\large \ 
}$z_{0}=0,${\large \ }$\Omega =1,$ and $\kappa =0.8$. To find the value of
the parameter $\lambda $, we have considered the case of lower\ sign in Eq. (%
\ref{13}). We note that the dark solitary wave solutions are the particular
cases of these gray solitary wave solutions when the constraint $\lambda
=B/(1+D)$ is satisfied.

\begin{description}
\item[\textbf{3. W-shaped solitary waves}] 
\end{description}

We have also obtained two W-shaped solitary wave solutions for Eq. (\ref{1}%
). The first case takes place when the conditions $\lambda >0$ and $\lambda
-B/(1+D)<0$ are satisfied. The W-shaped solitary wave solution in this case
has the form, 
\begin{equation}
E(z,x)=\left\vert \lambda -\frac{B}{1+D\cosh (w(x-q(z-z_{0})))}\right\vert
\exp [i\left( \kappa z-\Omega x+\theta \right) ],  \label{17}
\end{equation}%
where the parameters $\lambda $, $w$, $B$ and $D$ are given by Eqs. (\ref{13}%
-\ref{15}).

The second case takes place when the following conditions $\lambda <0$ and $%
\lambda -B/(1+D)>0$ are satisfied. The W-shaped solitary wave solution in
this case has the form, 
\begin{equation}
E(z,x)=-\left\vert \lambda -\frac{B}{1+D\cosh (w(x-q(z-z_{0})))}\right\vert
\exp [i\left( \kappa z-\Omega x+\theta \right) ],  \label{18}
\end{equation}%
where the parameters $\lambda $, $w$, $B$ and $D$ are given by Eqs. (\ref{13}%
-\ref{15}) with the replacement of parameter $b$ to $-b$.

Figure 1(c) displays the propagation of the solitary wave solution (\ref{17}%
) for the parameter values{\large \ }$\alpha =2,${\large \ }$\sigma =\frac{1%
}{15},${\large \ }$\nu =\frac{1}{45},${\large \ }$q=0.1,${\large \ }$%
z_{0}=0, ${\large \ }$\Omega =\frac{1}{3},$ and $\kappa =0.2$. To find the
value of the parameter $\lambda $, we have considered the case of lower\
sign in Eq. (\ref{13}). One can see from this figure that the structure
takes the shape of W, which can be formed in the waveguide medium due to a
balance among the diffraction and quadratic-cubic nonlinearities.

\begin{description}
\item[\textbf{4. Periodic waves}] 
\end{description}

We have also obtained an exact periodic wave solution for Eq. (\ref{5}) as 
\begin{equation}
u(\xi )=\frac{A}{B+\cos (w(\xi -\xi _{0}))},  \label{19}
\end{equation}%
where the real parameters $w$ and $B$ are 
\begin{equation}
w=\sqrt{c},\qquad B=\frac{\pm 1}{\sqrt{1-Q}},\qquad Q=\frac{9ac}{2b^{2}},
\label{20}
\end{equation}%
with $c>0$ and $Q<1$. The real parameter $A$ is given by 
\begin{equation}
A=\mp \frac{3c}{b\sqrt{1-Q}}.  \label{21}
\end{equation}%
The periodic wave in Eq. (\ref{19}) is a bounded solution for the condition $%
|B|>1$, which yields $Q>0$. Hence, we have the following conditions for the
bounded periodic solution given in Eq. (\ref{19}): 
\begin{equation}
c>0,\qquad 0<\frac{9ac}{2b^{2}}<1.  \label{22}
\end{equation}%
The exact periodic bounded wave solution of Eq. (\ref{1}) has the form,%
\begin{equation}
E(z,x)=\frac{A}{B+\cos (w(x-q(z-z_{0})))}\exp [i\left( \kappa z-\Omega
x+\theta \right) ].  \label{23}
\end{equation}

An example of propagation of the nonlinear wave solution (\ref{23}) is shown
in Fig. 1(d) for the parameter values $\alpha =1,$ $\sigma =0.25,$ $\nu
=-0.5,$ $q=0.1,$ $z_{0}=0,$ $\kappa =\frac{25}{48}$ and $\Omega =1$. It is
interesting to see that this structure presents an oscillating behaviour
superimposed at a nonzero background.

\begin{figure}[h]
\includegraphics[width=1\textwidth]{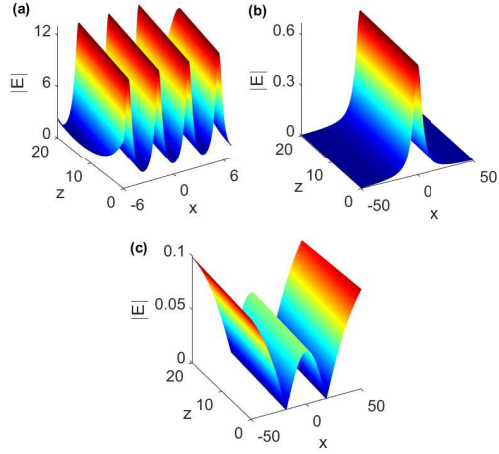}
\caption{Evolution of nonlinear wave solutions (a) modified periodic
wave (\ref{29}) with parameters $\alpha =-1,$\ $\sigma =0.5,$\ $\nu =0.5,$\ $%
q=0.1, $\ $z_{0}=0,$\ $\Omega =1,$\ and $\kappa =0.5$ (b) bright algebraic
solitary wave (\ref{34}) with parameters $\alpha =1,$ $\sigma =0.25,$ $\nu =-0.5,$%
\ $z_{0}=0,$ $\Omega =2,$ $q=0.1$, and $\kappa =2$ (c) W-shaped algebraic
solitary wave (\ref{38}) with parameters $\alpha =-1,$ $\sigma =0.03,$ $\nu =0.1,$ 
$\kappa =0.00045,$ $q=0.2,$ $\Omega =0.06,$ and $z_{0}=0$.}
\label{FIG.2.}
\end{figure}

\begin{description}
\item[\textbf{5. Modified periodic waves}] 
\end{description}

We have also obtained modified periodic wave solution for Eq. (\ref{5}) as
follows:%
\begin{equation}
u(\xi )=\left\vert \lambda -\frac{A}{B+\cos (w(\xi -\xi _{0}))}\right\vert ,
\label{24}
\end{equation}%
where $\lambda \neq 0$. In this case the parameter $\lambda $ is given by
equation $a\lambda ^{2}+b\lambda +c=0$. The real parameters $\lambda $, $w$
and $B$ are 
\begin{equation}
\lambda =\frac{1}{2a}\left( -b\pm \sqrt{b^{2}-4ac}\right) ,\qquad w=\sqrt{%
-2c-b\lambda },  \label{25}
\end{equation}

\begin{equation}
B=\frac{\pm 1}{\sqrt{1+R}},\qquad R=\frac{9a(2c+b\lambda )}{2(b+3a\lambda
)^{2}},  \label{26}
\end{equation}%
where $b^{2}>4ac$, $2c+b\lambda <0$ and $R>-1$. Also, the parameter $A$ is
given by 
\begin{equation}
A=\mp \frac{3(2c+b\lambda )}{(b+3a\lambda )\sqrt{1+R}}.  \label{27}
\end{equation}%
Another solution has the form, 
\begin{equation}
u(\xi )=-\left\vert \lambda -\frac{A}{B+\cos (w(\xi -\xi _{0}))}\right\vert ,
\label{28}
\end{equation}%
where $\lambda \neq 0$. The parameters in this solution follow from Eqs. (%
\ref{25}-\ref{27}) with the replacement of parameter $b$ to $-b$. Thus the
appropriate modified periodic bounded solutions of Eq. (\ref{1}) are 
\begin{equation}
E(z,x)=\pm \left\vert \lambda -\frac{A}{B+\cos (w(x-q(z-z_{0})))}\right\vert
\exp [i\left( \kappa z-\Omega x+\theta \right) ].  \label{29}
\end{equation}

In Fig. 2(a), we have presented an example of propagation of the nonlinear
wave solution (\ref{29}) for the parameter values $\alpha =-1,$\ $\sigma
=0.5,$\ $\nu =0.5,$\ $q=0.1,$\ $z_{0}=0,$\ $\Omega =1,$\ and $\kappa =0.5$.
To get the value of the parameter $\lambda $, we have considered the case of
upper\ sign in Eq. (\ref{25}). It is clear from the figure that the profile
of the wave presents the periodic property as it propagates inside the
waveguide.

\begin{description}
\item[\textbf{6. Bright algebraic solitary waves}] 
\end{description}

The nonlinear differential equation (\ref{5}) supports the exact
algebraic-type solitary wave solution:%
\begin{equation}
u(\xi )=\frac{p}{1+\mu (\xi -\xi _{0})^{2}}.  \label{30}
\end{equation}%
There are two different cases for parameters in Eq. (\ref{30}). In the first
case [$u(\xi )\geq 0$] the real parameters $p$ and $\mu $ are defined by the
expressions, 
\begin{equation}
p=-\frac{4b}{3a},\qquad \mu =\frac{2b^{2}}{9a},  \label{31}
\end{equation}%
and $c=0$ in Eq. (\ref{6}). Thus the wave number $\kappa $ of this optical
wave solution is given by%
\begin{equation}
\kappa =\frac{1}{2}\alpha \Omega ^{2}.  \label{32}
\end{equation}%
In this case [$u(\xi )\geq 0$] we have the conditions $p>0$ and $\mu >0$
which yield $a>0$ and $b<0$.

In the second case [$u(\xi )\leq 0$] the real parameters $p$ and $\mu $ are
defined by the expressions, 
\begin{equation}
p=\frac{4b}{3a},\qquad \mu =\frac{2b^{2}}{9a},  \label{33}
\end{equation}%
and $c=0$ in Eq. (\ref{6}). We replaced here $b$ to $-b$ which follows from
relations in Eq. (\ref{7}). Thus in this second case we have the conditions $%
p<0$ and $\mu >0$ which yield $a>0$ and $b<0$, and the wave number $\kappa $
is given by (\ref{32}). Combining Eqs. (\ref{2}) and (\ref{30}) we have an
exact algebraic solitary wave solution to\ the quadratic-cubic NL equation (%
\ref{1}) of the form, 
\begin{equation}
E(z,x)=\frac{p}{1+\mu \lbrack x-q(z-z_{0})]^{2}}\exp [i\left( \kappa
z-\Omega x+\theta \right) ].  \label{34}
\end{equation}

Figure 2(b){\large \ }presents an example of the evolution of the algebraic
solitary wave solution (\ref{34}) for the parameter values $\alpha =1,$ $%
\sigma =0.25,$ $\nu =-0.5,$\ $z_{0}=0,$ $\Omega =2,$ $q=0.1$, and $\kappa =2$%
. Clearly, the wave profile take a bright localized structure on a zero
background.

\begin{description}
\item[\textbf{\ 7. W-shaped algebraic solitary waves}] 
\end{description}

We have obtained another exact algebraic solitary wave solutions for Eq. (%
\ref{5}) as follows:%
\begin{equation}
u(\xi )=\pm \left\vert \lambda -\frac{p}{1+\mu (\xi -\xi _{0})^{2}}%
\right\vert .  \label{35}
\end{equation}%
We have here two different cases: (1) with $u(\xi )>0$, and (2) with $u(\xi
)<0$. We take in (\ref{35}) the signs $(+)$ and $(-)$ for the first and
second case respectively. Note that we consider here $\lambda \neq 0$
because the case with $\lambda =0$ is given in Eq. (\ref{30}). In the first
case real parameters $p$, $\mu $ and $\lambda $ are given by%
\begin{equation}
p=-\frac{2b}{3a},\quad \mu =\frac{b^{2}}{18a},\quad \lambda =-\frac{b}{2a}.
\label{36}
\end{equation}%
We also have the relation $c=b^{2}/4a$ for parameter in Eq. (\ref{6}). Thus
the wave number $\kappa $ for this solution is given by%
\begin{equation}
\kappa =\frac{\alpha b^{2}}{8a}+\frac{\alpha }{2}\Omega ^{2}.  \label{37}
\end{equation}%
We note that derivative $du(\xi )/d\xi $ of the W-shaped solution is not a
continuous function at two points $\xi =\xi _{0}\pm 1/\sqrt{3\mu }$ where
the function $u(\xi )$ is equal to zero. Hence, in the first case we have $%
\mu >0$ and $a>0$ for the W-shaped solution.

In the second case [$u(\xi )<0$] we take the sign $(-)$ in Eq. (\ref{35})
and also replace the parameter $b$ to $-b$ in Eq. (\ref{36}). This change is
connected with relations presented in Eq. (\ref{7}). In the second case we
also have $\mu >0$ and $a>0$ for the W-shaped solution, and the wave number $%
\kappa $ is given by Eq. (\ref{37}). Further substitution of the solutions (%
\ref{35}) into Eq. (\ref{2}) yields an exact algebraic solitary wave
solutions of Eq. (\ref{1}) for two cases with an appropriate signs as 
\begin{equation}
E(z,x)=\pm \left\vert \lambda -\frac{p}{1+\mu \lbrack x-q(z-z_{0})]^{2}}%
\right\vert \exp [i\left( \kappa z-\Omega x+\theta \right) ].  \label{38}
\end{equation}

An example of the evolution of the algebraic solitary wave solution (\ref{38}%
) is shown in Fig. 2(c) for the parameter values $\alpha =-1,$ $\sigma
=0.03, $ $\nu =0.1,$ $\kappa =0.00045,$ $q=0.2,$ $\Omega =0.06,$ and $%
z_{0}=0 $. It is interesting to see that this nonlinear waveform is a
W-shaped algebraic solitary wave.

\section{Conclusion}

To conclude, we have studied the transmission dynamics of light beams
through a non-centrosymmetric waveguide exhibiting second- and third-order
nonlinearities. We have presented new types of exact analytical localized
and periodic wave solutions for the quadratic-cubic NLS equation that can
model the propagation of optical beams in such system. The newly found
solutions include gray and W-shaped solitary waves as well as periodic wave
solutions. We have also obtained the exact algebraic bright and W-shaped
solitary wave solutions of the model. No doubt, the derived structures may
be helpful in understanding the physical phenomena in dynamical systems with
quadratic-cubic nonlinearities.

\end{document}